# Interpreting Compton anisotropy of ice $I_h$: a cluster partitioning method

Sébastien RAGOT[#], Jean-Michel GILLET[#], Pierre J. BECKER[#@]

[#] Laboratoire Structure, Propriété et Modélisation des Solides (CNRS, Unité Mixte de Recherche 85-80). École Centrale Paris, Grande Voie des Vignes, 92295 CHATENAY-MALABRY, FRANCE

[@] Université Catholique de l'ouest. 1, place André-Leroy. BP808, 49008 ANGERS Cedex 01, FRANCE

## Abstract

We propose a simple cluster-based method with application to calculations of Compton profile anisotropies of ice. The convergence of the method is checked with respect to Crystal95 results. Increasing both basis-set quality and cluster sizes results in a decrease of the magnitude of theoretical Compton anisotropies. The agreement with experimental data is therefore improved towards previously calculated anisotropies. Moreover, analyzing directional autocorrelation functions shows an evidence for both anti-bonding and polarization effects.

# I. Introduction

The complexity of interactions between water molecules fascinates scientists since the 30s. A major question that still remains unanswered concerns the inability of ice $I_h$ to form a totally ordered structure even at the lowest temperatures[1]. Besides, the astonishing ability for ice to crystallize in many different forms depending on pressure and temperature results from peculiar properties of hydrogen bonds. Hydrogen bonds are still widely studied[2,3,4], partly due to possibilities offered by diffraction experiments[5]. Moreover, new theoretical approaches have recently been developped[2] and the possibilities offered by topological analyses of periodic systems (program TOPOND[6]) open the field of new ways of characterizing hydrogen bonds[7]. Beside diffraction, incoherent scattering experiments reveal details about dynamics of either nuclei (for example through inelastic neutron scattering[8]) or electrons (through Compton scattering). Here we focus on possibilities offered by Compton scattering measurements for bonding effect studies between water molecules. In 1999, Isaacs et al.[19] measured for the first time directional Compton profiles of ice $I_h$, whereas former experiments[9,10,11] focused on isotropic Compton profile measurements. Before commenting on the interpretation the authors gave to their measurements, we remind general features of ice $I_h$.

Local molecular arrangements usually proposed for ice $I_h$ are assumed to follow the Bernal-Fowler[12] rules whereas the average structure of ice relies on the statistical model of Pauling[13]. The average structure of ice was understood in 1953 by Owsten[14] and later confirmed by Peterson and Levy[15]. Both have accredited the statistical model of Pauling. The average structure belongs to $P6_3/mmc$ centro-symmetric space group. Two sites are available for the protons (*i.e.* H1 and H2 on figure 1.a). Together with an O atom, H1 forms an hydrogen bond along the *z*-axis, whereas H2 forms a bond along a direction close to a plane perpendicular to *z*. However, ice-rules allow for only one crystallographic sub-structure at a time, each belonging to space-group $P2_1$ (figure 1.b). The O-O distance in ice[18] is about 2.75

Å and must be compared to the average distance of 2.98 Å observed for the gas phase dimer[16]. Since X-ray diffraction is a bulk probe, sensitive to a macroscopic average of the sample, it is not possible to extract information inherent to a locally ordered domain. Van Beek[17] has proposed a superimposition of 6 locally ordered domains whose structure obeys the ice-rules. Each substructure gives rise to $P2_1$ symmetry, with $c$ (along direction $z$, see figure 1) as a unique axis. There are, in each case, 2 crystallographically unique molecules[18]. Van Beek has calculated the structure factors of ice $I_h$ by averaging the *ab initio* structure factors corresponding to each configuration, resulting in a good agreement with their experimentally determined counterparts[17].

As mentioned above, accurate directional Compton profiles of ice $I_h$ have been measured by Isaacs *et al.*[19], allowing for a comparison between experimental and *ab initio* Compton anisotropies (CAs). CAs are usually very sensitive to the 2-center terms of the 1RDM[20,21]. The published anisotropy (figure 3) is the difference between directional Compton profiles measured along the $z$-axis (O-H1-O direction, see figure 1.b) and an average direction perpendicular to $z$ (later denoted by $x/y$), so that the CA refers to $J_z(q) - J_{x/y}(q)$. The authors interpreted the oscillations of the measured CA as a direct proof for covalence between water molecules in ice. Similar oscillations were obtained from a density-functional theory (DFT) calculation (figure 3), where molecules were oriented according to the Bernal-Fowler ice rules. However, the magnitude of the theoretical CA had to be reduced of 40% in order to match the experimental curve. A recent DFT-based investigation of Romero and co-workers has further led to the same conclusion[22]. On the other hand, superimposing signals pertaining to fully independent molecules does not reproduce the oscillations[19]. Finally, the power spectrum of the CA, *i.e.* the square modulus of the Fourier transform of the CA, exhibits 2 main peaks located at positions evoking both intermolecular H⋯O et O⋯O distances. The conclusion of Isaacs *et al.*[19] gave rise to a vivid controversy and was notably discussed by Ghanty et al.[23] who concluded: *"the oscillations are irrelevant to the discussion of the covalent character of the bond. Rather they just reflect the result of antisymmetrizing the product of monomer wave functions »*. Conclusions of Ghanty *et al.* issued from considerations on an ice-like dimer $(H_2O)_2$, the CA of which again overestimates the magnitude of the experimental CA of ice by a factor of about 2.

Finally, many points remain to be clarified. Why do the theoretical magnitudes of CA overestimate the experimental ones? What is the effect of bonding/anti-bonding interactions on the anisotropies? How can we interpret the presence of 2 main peaks on the PS?

In this paper, we apply a cluster partitioning method (CPM, section II) to ice in order to compare the effects of both basis-sets and interactions of increasing range on the CA (section III). We further discuss the effects of anti-bonding interactions on the autocorrelation function in the last section.

## II. A cluster partitioning Method[24]

In the case of an isolated molecule, treated within the Born-Oppenheimer (BO) approximation, the first order reduced density matrix integrated over spin variables (1RDM) writes as $\rho(r,r') = \sum_{A,B} \sum_{i \in A, j \in B} c_{ij}^{AB} \varphi_{iA}(r) \varphi_{jB}^*(r')$. Atomic orbitals $\varphi_{iA}$ are centered on $R_A$ (pointing at the center of an atom) and are assumed to be real, for simplicity. Note that the 1RDM can formally be rewritten as

$$\rho(r,r') = \sum_{A,B} \rho_{A,B}(r - R_A, r' - R_B) \qquad (1)$$

Separating 1- and 2-center terms in (1) and symmetrizing them afterwards yields the following decomposition of the 1RDM

$$\rho = \sum_A \left\{ \rho_{A,A} + \tfrac{1}{2} \sum_{B(\neq A)} [\rho_{A,B} + \rho_{B,A}] \right\} \qquad (2)$$

This partition scheme is, so far, nothing else than a Mulliken-like partition scheme, which allows for rewriting the 1RDM as $\rho(r,r') = \sum_A \rho_A(r - R_A, r' - R_A)$ or conversely

$$\tilde{\rho}(R,s) = \sum_A \tilde{\rho}_A(R - R_A, s) \qquad (3)$$

$\tilde{\rho}$ refers to the intracular-extracular representation of the 1RDM[25], $R$ stands for $(r + r')/2$, whereas $s$ is the difference vector $r - r'$. Other partition schemes could obviously result in a 1-center decomposition of the 1RDM similar to (3).

When extending the molecule to a crystal with a group of $N$ atoms as a unit basis, $\tilde{\rho}(R,s)$ becomes invariant by a translation $L$ (a lattice vector) of the $R$ coordinate

$$\tilde{\rho}(R,s) = \sum_L \sum_{A=1}^N \tilde{\rho}_A(R - L - R_A, s) \equiv \sum_L \tilde{\rho}_L(R - L, s) \qquad (4)$$

The momentum density is defined as

$$n(p) = \frac{1}{(2\pi)^3} \int \rho(r,r') e^{ip.(r-r')} dr dr' = \frac{1}{(2\pi)^3} \int \tilde{\rho}(R,s) e^{ip.s} dR ds \qquad (5)$$

$n(p)$ turns out to be the Fourier transform of the so-called autocorrelation function[20], which is a position-space function obtained after integration over the $R$ coordinate of the 1RDM.

$$B(s) = \int \tilde{\rho}(R,s) dR \qquad (6)$$

Finally, each basis unit term $\tilde{\rho}_L(R - L, s)$ from (4) gives an identical contribution to $n(p)$ and $B(s)$, so that the arbitrariness of the 1RDM partition disappears and only one term is to be computed. The directional impulse Compton profile[20] $J_z(q)$ is then obtained[26]:

$$J_z(q) = \int n(p)\delta(p_z - q)dp = \frac{1}{2\pi} \int B_z(s_z) e^{iqs_z} ds_z \qquad (7)$$

Practically, two-center contributions in (2) vanish as overlaps between orbitals become negligible. For example, in the case of NaCl-like crystals, calculations can be performed on 2 clusters, respectively centered on Na and Cl ions, in order to preserve the local symmetry of the environment. Electronic densities can then be recovered by summing the contributions $\widetilde{\rho}_{Na}(\boldsymbol{R},\boldsymbol{s})$ and $\widetilde{\rho}_{Cl}(\boldsymbol{R},\boldsymbol{s})$, pertaining to each ion located *at the center* of their corresponding cluster. For finite cluster calculations, this method is obviously only approximate but we have checked that 1-electron properties converge rather quickly towards crystal ones, as calculated with the Crystal95 program package (Crystal95)[27] at Hartree-Fock (HF) level, in the case of insulators and semiconductors. We expect one-electron properties to converge even faster for molecular crystals.

Even approximate, the Cluster Partitioning Method (CPM) should provide some advantages. As a molecular-like approach, it bypasses the summation over the first Brillouin zone (no periodicity of the system is needed). As such, it also permits investigations of defects, provided that an appropriate partition scheme for the 1RDM is available. Moreover, using electronic structure calculation codes such as GAUSSIAN94 (G94)[28] further allows for calculations at correlated level with explicitly correlated wave functions (applications to insulators are currently investigated).

## III. Theoretical *vs.* experimental Compton anisotropies

As stated above, the CPM can be extended to molecules in a crystal. In the following, the 1RDM partition refers to the crystallographically independent water molecules.

### *III.1 Choice of clusters and basis-sets*

The clusters we selected mimic the $P2_1$ symmetry of the infinite crystal around the 2 crystallographically independent molecules and further obey the Bernal-Fowler rules. Two different sizes are chosen: clusters contain either 22 or 46 molecules (hereafter denoted by c22 and c46, respectively). Calculations have been performed within STO-3G, double zeta (DZ) and double zeta + polarization (DZP) Basis-sets[29,30]. In order to appreciate influences of larger basis-sets (like cc-pVTZ[30]), we also performed calculations on a cluster of 8 molecules (2 central molecules + first neighbors for the active space = 8 molecules), surrounded by 51 molecules simulated by point charges. This cluster will be later referred to as c8(59). Subsequent errors on the total number of electrons are negligible (typically 0.01-0.1 %).

*III.2 Convergence of CA*

As already mentioned, the theoretical (DFT) magnitude of the CA[31] overestimates the experimental one by a factor 1.7. Possible reasons evoked by Isaacs *et al.* are thermal effects, zero-point vibrations, electronic correlation and/or disorder. The recent investigation of Romero *et al.*[22] suggests that finite temperature effects can be ruled out as one of the possible causes of discrepancy between theory and experiment. Here we compare the influences on CA of both the cluster size (so the long-range interactions) and the basis-set. Calculations are performed at HF level.

First, we compare on figure 2.a cluster anisotropies issued from c22, c46 and Crystal95, calculated within STO-3G basis-set. The cluster (c46/STO-3G) and Crystal95 anisotropies are quasi identical. We have also checked the convergence on full directional Compton profiles (not reported): relative difference between c46/STO-3G and Crystal95 do not exceed 0.3% for $q < 2$ u.a. The convergence is thus considered to be reached. As expected, increasing the cluster size decreases the magnitude of the CA. Figure 2.b shows that the CA of the ice-like dimer is about twice the CA of c46 in magnitude (for $q < 1.5$ u.a.), when using a DZP basis-set. The c46/DZP anisotropy has a lower magnitude than the c46/STO-3G one, so that increasing the basis-set quality shall also contribute to decrease the magnitude of the CA. Note that the larger basis-set (DZP) gives rise to a small oscillation at low $q$ (for $q < 0.6$ a.u., see fig.2.b).

Using smaller clusters allows for calculations with higher basis-set quality. The tests we performed at HF/cc-pVTZ level with c8(59) clusters have suggested that convergence is not fully reached at HF/DZP level[32].

Besides, one advantage provided by molecular quantum methods is the possibility to explicitly take electron correlation into account. Usually, correlation effects are mostly isotropic on momentum distributions[21] (as indicated by a current work on insulators), so that consequences on CA are weak. For instance, tests have been done at MP2/cc-pVTZ and QCISD/DZP level for the ice-like dimer: subsequent results have shown that correlation brings changes less than 0.5% on the Compton anisotropy. As a consequence, even if the basis-sets chosen underestimate correlation effects, we think that correlation is not crucial for the interpretation of the measured CA.

*III.3 Comparison with experimental results*

We now compare the CA issued from c46/DZP calculations with both DFT and experimental results of Isaacs *et al.*[19]. In each case, theoretical CAs are scaled in order to match the main experimental peak. One notices first that all calculations result in a good phase agreement with the experimental CA (figure 3). This supports the Bernal-fowler model regarding the momentum space, *i.e.* the off-diagonal part of the 1RDM. Note that in position space, it is necessary to mix different $P2_1$ substructures in order to reproduce the experimental charge density[17]. In momentum space: any $P2_1$ substructure leads to quasi-identical CAs, with a good agreement with the experimental one. Figure 3 also shows that the CPM provides better qualitative agreement with experimental CA than the DFT-based calculation, as obtained by scanning the published figure of reference 19. Moreover, scaling factors used for the CPM results are closer to 1 (see caption). However, notice that it is not the purpose of DFT to be accurate in momentum space. As quoted by Pathak et al.[33]: « While the KS scheme yields, in principle, the exact coordinate-space electron density, it does not necessarily give the correct momentum density (…) as pointed out by Lam and Platzman[34] ».

## IV. Analysis of theoretical autocorrelation functions

It is possible to analyze the anisotropy in position space through the transformation:

$$B_z(s) - B_{x/y}(s) = 2\pi \int_{-\infty}^{\infty} [J_z(q) - J_{x/y}(q)] e^{iqs} dq \qquad (8)$$

The so-called power spectra[19] are defined as $|B_z(s) - B_{x/y}(s)|^2$: they are compared on figure 4. Again, a good agreement is observed, regardless of the magnitudes. The present experimental power spectrum slightly differs from the original one[19] because integral (8) was recomputed in order to minimize artifacts due to truncation error[35]. As already mentioned, positions of the 2 main peaks (1.7 and 2.8 Å) evoke intermolecular distances H⋯O (hydrogen bonds) and O⋯O of respectively 1.75 and 2.75 Å.

### *IV.1 Interpretation of* $B_z(s) - B_{x/y}(s)$

Following Isaacs *et al.*, we now compare the differences $J_z(q) - J_{x/y}(q)$ and $B_z(s) - B_{x/y}(s)$ obtained from (a) the CPM, (b) a model that consists of fully independent molecules and (c) a model where independent molecules are surrounded by point charges that simulate electrostatic influence of neighboring atoms. Calculations are performed at HF/DZP level. As stated by Isaacs *et al.*, the anisotropy obtained from a fully independent molecule model (IM on figure 5) does clearly not reproduce the characteristic oscillations. Taking into account purely electrostatic effects (IM+C, figure 5) leaves the conclusion unchanged. Therefore, one has to consider other mechanisms. Rather than an anisotropy, we now focus on directional autocorrelation functions. We compare on figure 6 the autocorrelation functions provided by either the independent molecule model or the CPM. In each case, autocorrelation functions take negative values, a fingerprint of possible polarization or anti-bonding effects[36]. In independent molecule model, the autocorrelation functions take negative values because of polarization effects only. Oscillations observed on CPM curves denote *additional* anti-bonding interactions, *i.e.* negative coefficients in the 1RDM, which change the sign of the autocorrelation function near 2.8 Å. These oscillations are obviously more pronounced in the *z*-direction (see figure 1) than in the *x*-direction. The anti-bonding character can be thought of as resulting from interactions between closed-shell systems (H2O molecules) or even, between paired electrons of one intra-molecular O-H bond and the adjacent oxygen doublet[37]. As mentionned by Ghanty et al.[23]: whether the anti-bonding character of interactions is compatible with the definition of covalence is merely a question of terminology, though unusual. Notice that the anti-bonding character of interactions is dominant in momentum space in spite of the fact that the system remains globally "bonded", notably through dipole-dipole interactions. However, it is necessary to point out that a purely antisymmetrized product of isolated monomer wave functions[23] does not correctly describe experimental features of the CA at small momenta, even qualitatively. This can be checked by comparing corresponding results of Ghanty et al.[23] with the experimental CA at $q < 0.6$ a.u. [38] on figure 3. Neither minimal basis-set nor dimer-based calculations do satisfactorily describe the low-$q$ oscillation (fig. 2.a and 2.b). Conversely, bulk-DFT and c22 or c46/DZP calculations result in a good agreement (fig. 3). Therefore, the low-$q$ oscillation can be interpreted as another consequence of bulk interactions.

In order to interpret the 2 main peaks on the *power spectrum*, we followed the suggestion of Ghanty et al.[23], i.e. we replaced $H_2O$ molecules by Ne atoms, located at positions of O atoms. The resulting anisotropy of autocorrelation function (figure 7) is obviously smaller in magnitude (due to the lower polarizability of Ne atoms) but exhibits one negative and one positive peak at about 2.2 and 2.8 Å, respectively. The first peak can clearly not be ascribed to any kind of electron sharing between Ne atoms, since these are located at positions of O atoms. Rather, these peaks reflects the fact that anti-bonding effects are more pronounced along a Ne-Ne segment and that we considered a difference between 2 directional functions. Similarly, the peak at 1.7 Å in ice has no straightforward meaning.

# V. Conclusions

Cluster-based calculations, followed by a simple partition of the 1-electron reduced density matrix permits to take advantage of molecular methods for the estimation of the Compton profile anisotropy of ice. Within a minimal basis-set, the cluster partitioning method converge towards Crystal95 ones. Increasing either the basis-set quality or the cluster-size results in a decrease of Compton anisotropy magnitudes: long-range interactions reinforce the isotropy of momentum distribution. It could therefore be interesting to analyze how cooperative effects in ice can be correlated to the decrease of the magnitude of the Compton anisotropy. Comparison with a recent Compton scattering experiment results in an improved agreement between experimental and theoretical data. This agreement is attributed to the use of a large basis-set, which is not prohibitive for the cluster partitioning method. Moreover, an analysis of directional autocorrelation functions clearly reveals both anti-bonding and polarization effects, which affect the measured Compton anisotropy. Anti-bonding effects create oscillations on Compton profiles, the magnitudes of which are partly driven by polarization effects. A comparison between anisotropies of autocorrelation functions issued from "ice-like" lattices of $H_2O$ and Ne has further shown that the first peak at 1.7 Å is not a characteristic of electron sharing. As a consequence, Compton oscillations are certainly irrelevant to the discussion of the covalent character of the bond[23] but not to that of the global cohesion mechanism.

Concurrently, a first study on disordered clusters indicates that a significant occurrence of Bjerrum[39] defects ($L$ and $D$ defects) should result in a visible peak at small $s$ on the anisotropy of autocorrelation functions. Such a peak arises due to the electronic coupling between 2 adjacent H atoms located on an O⋯O segment, which violates ice-rules. Whether this feature is visible or not on the experimental curve is not clear. This is however not surprising since both kind of defects have a high energetic cost: the molecular rearrangements are probably more subtly correlated. In comparison, ionic defects do not significantly affect the anisotropy. Notice finally that molecular disorder (including proton disorder and vibrations) must increase the overall symmetry. In that respect, the remaining discrepancy in magnitude between experimental and CPM anisotropy could be partly attributed to disorder.

## Acknowledgments

We are pleased to express our deep gratitude to A. Shukla who provided us with experimental data on ice, and for a careful reading of the manuscript, his suggestions and interesting discussions on the many features of the experimental Compton anisotropy .

## Captions to figures

Figure 1. (a): Statistical configuration of 4 molecules in $P6_3/mmc$ space-group (b): example of one possible $P2_1$ configuration of ice $I_h$.

Figure 2: Comparison of various theoretical Compton anisotropies. (a): convergence of CPM CAs at HF/STO-3G level. (b): Comparison of BS and cluster size effect on CA magnitude.

Figure 3: Comparison of experimental and theoretical CAs. Theoretical CAs are corrected for experimental resolution (0.14 u.a.). *Dots* : Experiment. *Full line* : c46/DZP (Scaling factor: 0.8). *Grey*: DFT (scaling factor: 0.6).

Figure 4: Comparison of power spectra corrected for experimental resolution. *Dots*: experiment. *Full line*: HF/c46/DZP. *Grey*: DFT results. Theoretical curves are individually scaled in order to match the experimental one.

Figure 5: Anisotropy $J_z$ - $J_{x/y}$ : Comparison between various models at HF/DZP level. *Black line*: cluster c46 results. *Grey*: fully independent molecule model (IM). *Dashed grey*: model of independent molecules surrounded by point charges (IM+C)

Figure 6: directional autocorrelation functions (ACF) in the 1.6 – 6.0 Å region. IM stands for independent molecule model.

Figure 7: Comparison of autocorrelation function (ACF) anisotropies of ice and Ne lattices (calculated at HF/cc-pVDZ level with c22 clusters ). The magnitude of Ne curve is multiplied by 20 for clarity.

# Figures

Figure 1.a

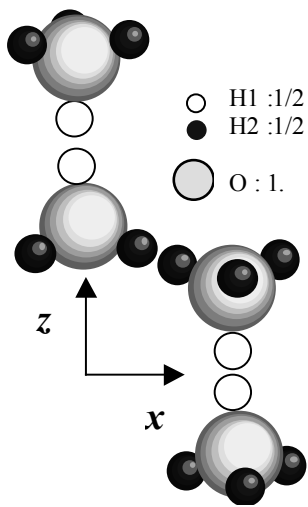

Figure 1.b

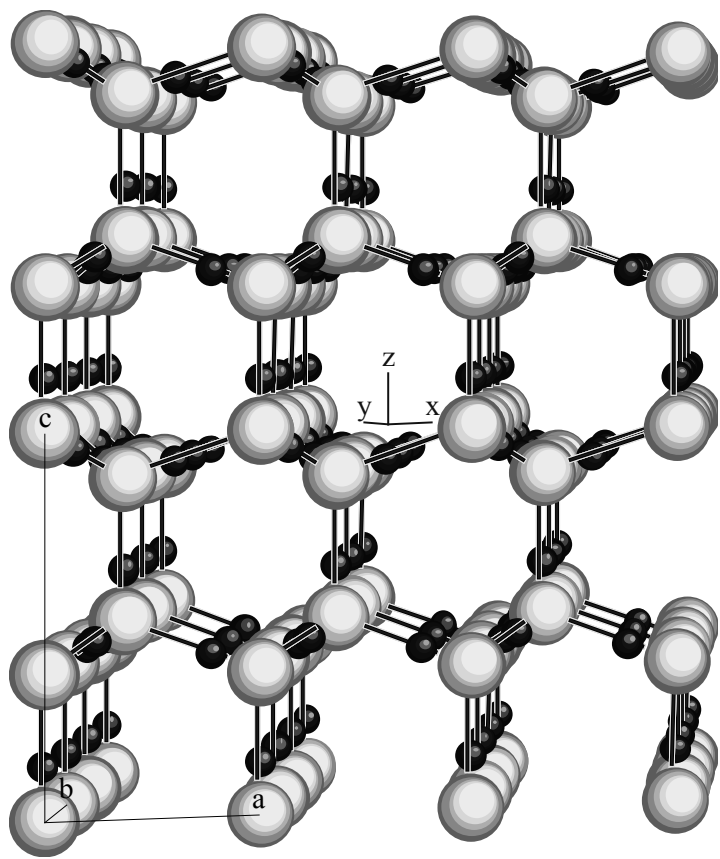

Figure 2.a

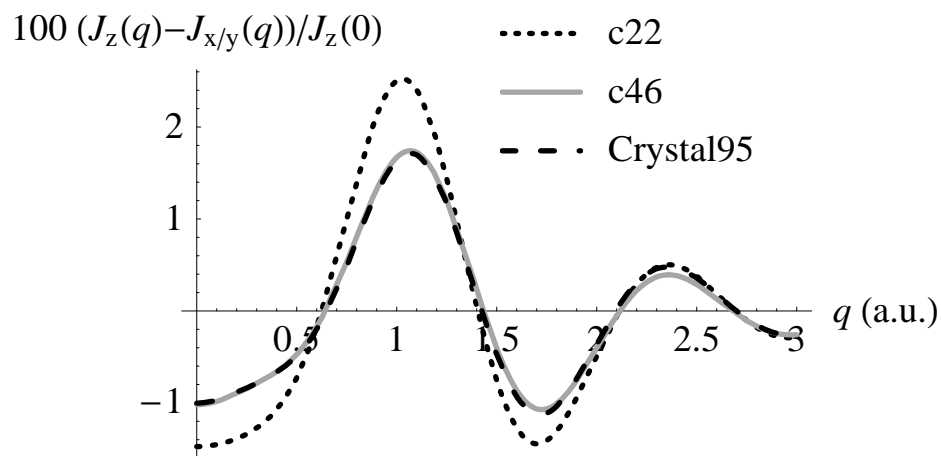

Figure 2.b

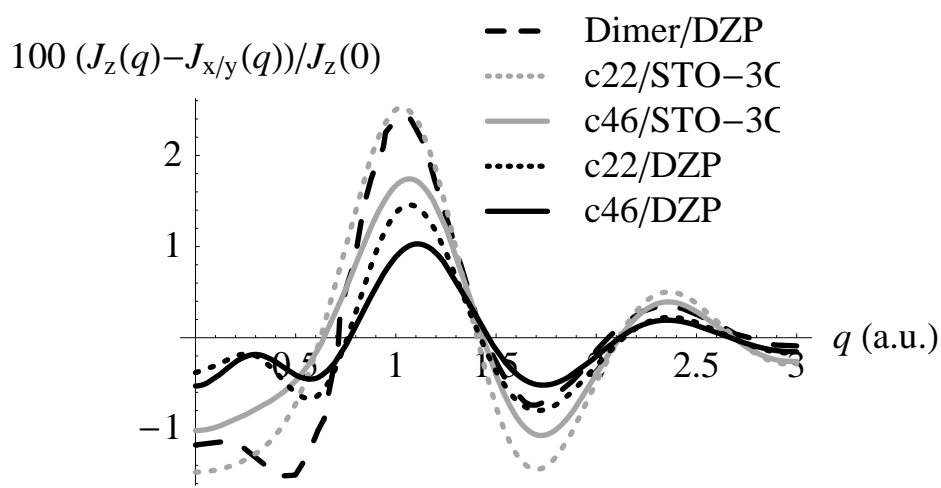

Figure 3

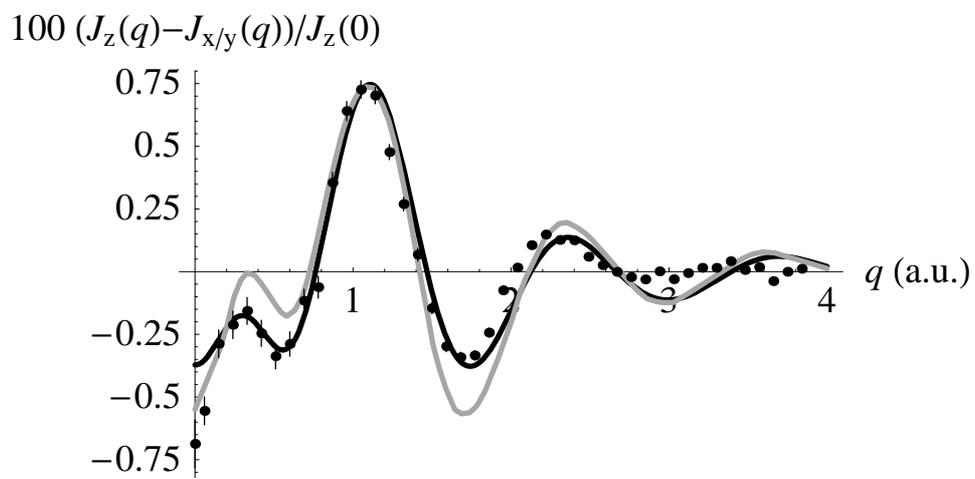

Figure 4

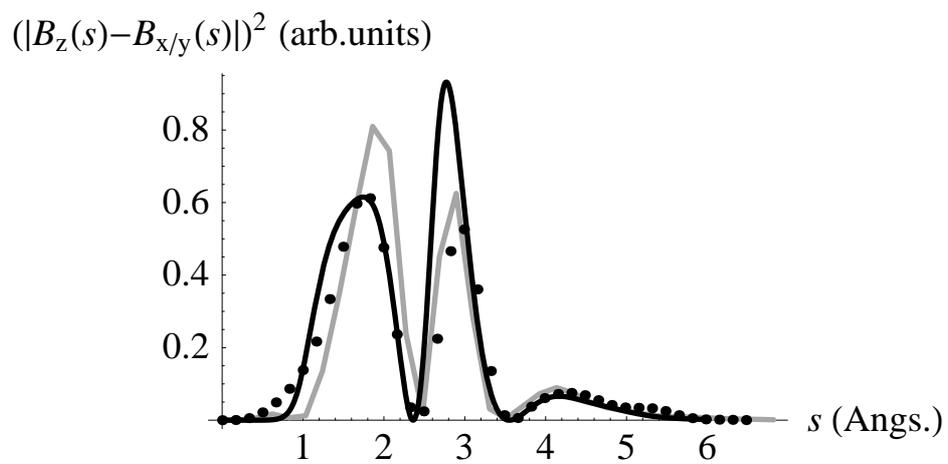

Figure 5

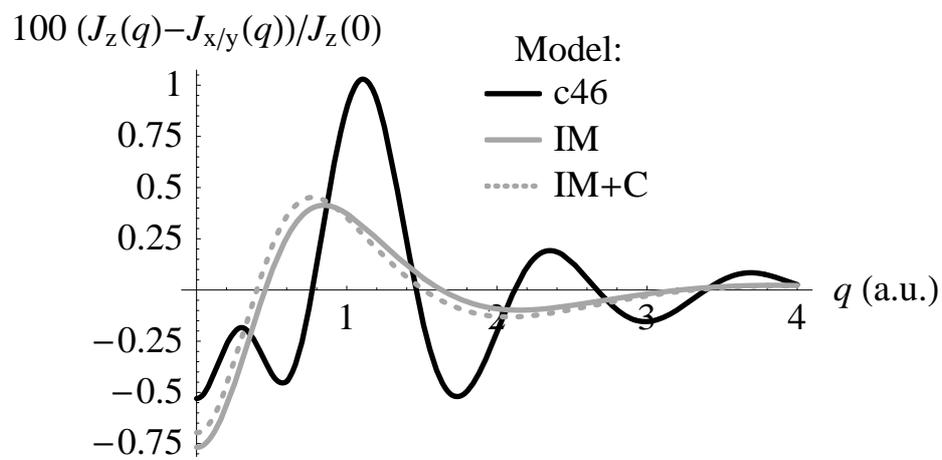

Figure 6

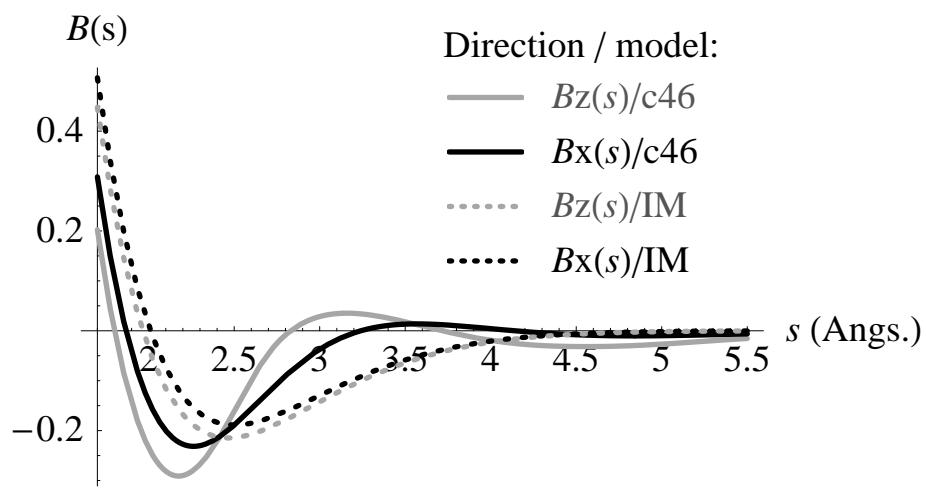

Figure 7

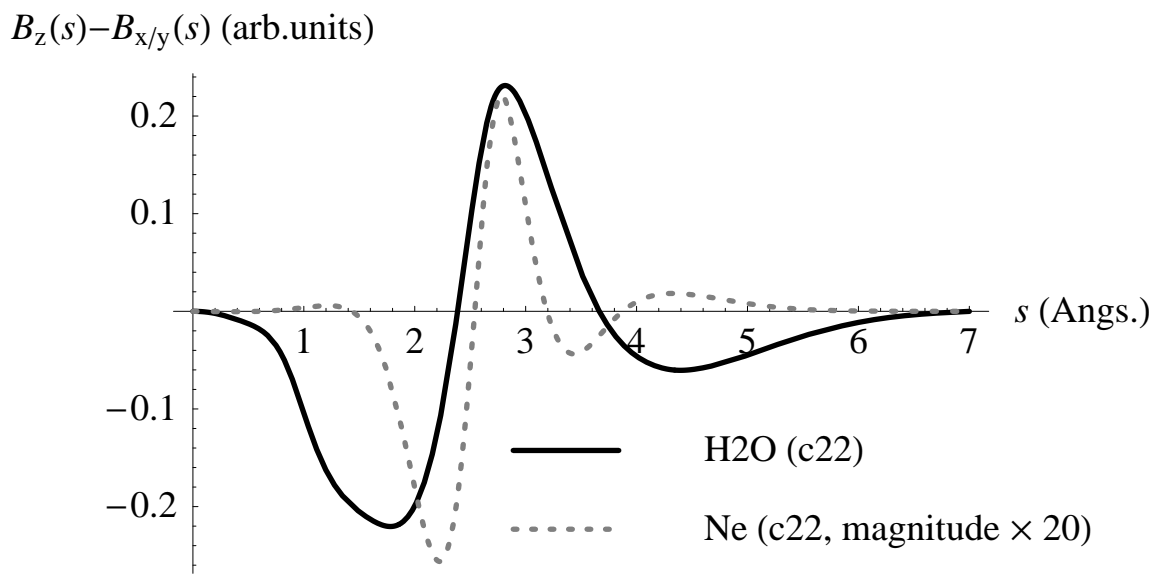

[26] According to Isaacs *et al*. (reference 19), we can neglect corrections to bring to this approximation in the case of the experiment on ice.

[27] R. Dovesi, V. R. Saunders, C. Roetti, M. Causà, N. M. Harrison, R. Orlando, and E. Aprà. *Crystal95 User's Manual*. University of Torino, Torino (1996).

[28] Gaussian 94, Revision D.4, M. J. Frisch, G. W. Trucks, H. B. Schlegel, P. M. W. Gill, B. G. Johnson, M. A. Robb, J. R. Cheeseman, T. Keith, G. A. Petersson, J. A. Montgomery, K. Raghavachari, M. A. Al-Laham, V. G. Zakrzewski, J. V. Ortiz, J. B. Foresman, J. Cioslowski, B. B. Stefanov, A. Nanayakkara, M. Challacombe, C. Y. Peng, P. Y. Ayala, W. Chen, M. W. Wong, J. L. Andres, E. S. Replogle, R. Gomperts, R. L. Martin, D. J. Fox, J. S. Binkley, D. J. Defrees, J. Baker, J. P. Stewart, M. Head-Gordon, C. Gonzalez, and J. A. Pople, Gaussian, Inc., Pittsburgh PA. (1995).

[29] Some of the basis sets used here were obtained from the Extensible Computational Chemistry Environment Basis Set Database. Contact David Feller or Karen Schuchardt for further information. E-mail: df_feller@pnl.gov. Web site: http://www.emsl.pnl.gov:2080/forms/basisform.html.

[30] For information on the STO-3G basis-set used, see W.J. Hehre, R.F. Stewart and J.A. Pople, J. Chem. Phys. **51**, 2657 (1969). For DZ and DZP, see T. H. Dunning, Jr., J. Chem. Phys. **53**, 2823 (1970); and T. H. Dunning, Jr. and P. J. Hay, in *Methods of Electronic Structure Theory*, Vol. 3, edited by H.F. Schaefer III (Plenum Press, 1977). For cc-pVDZ and cc-pVTZ, see T.H. Dunning, Jr., J. Chem. Phys. **90**, 1007 (1989).

[31] For details about the DFT calculation and bulk properties of ice $I_h$, see: D. R. Hamann. Phys. Rev. B **55**, R10157-R10160 (1997).

[32] The quality of basis-sets is critical for the outcome of calculations of Compton profiles as well as bulk properties; see for instance the work of: Ojamäe et al. for a description of various basis-sets tested for predicting properties of ice VIII. L. Ojamäe, K. Hermansson, R. Dovesi, C. Roetti, and V.R. Saunders, J. Chem. Phys **100**, 2128 (1994).

[33] R.K. Pathak, A. Kshirsagar, R. Hoffmeyer, and A. J. Thakkar, Phys. Rev. A **48**, 2946 (1974).

[34] L. Lam and P. M. Platzman, Phys. Rev. B **9**, 5122 (1974); **9**, 5128 (1974).

[35] A. Shukla, private communication.

[36] P. Patisson and W. Weyrich, J. Phys. Chem. Sol. **40**, 40 (1979).

[37] As indicated by a current investigation of 3-centers/4-electron models.

[38] A. Shukla, private communication. The experimental oscillation at $q < 0.6$ a.u. seems to be a characteristic of bulk long-range interactions. An investigation of this oscillation is in progress.

[39] N. Bjerrum, Kgl. Damske Videnskab. Selskab, Skr. **27**, 1 (1951); **27**, 41 (1951).